\documentstyle[12pt]{article}

\def\half{{\scriptstyle{1\over2}}}
\def\pifour{{\scriptstyle{\pi\over4}}}
\def\be{\begin{equation}}
\def\ee{\end{equation}}

\begin{document}
\begin{titlepage}
\title{Quantum properties of the polytopic action in some simple geometries}
\author{ {\bf
E.\ Alvarez} \\ Departamento de F\'\i sica Te\'orica, C XI \\
Universidad Aut\'onoma de Madrid\\ 28049 Madrid, Spain \and  {\bf J.\
C\'espedes} and {\bf E.\ Verdaguer} \\ Grup de F\'\i sica Te\`orica\\
Universitat Aut\`onoma de Barcelona\\ 08193 Bellaterra, Spain} \date{\null}
\maketitle
\begin{abstract}
The partition function corresponding to the ``polytopic" action, a new
action for the gravitational interaction which we have proposed recently, is
computed in the simplest two dimensional geometries of genus zero and
one. The functional integral over the Liouville field is approximated by an
ordinary integral over the constant zero mode. We study the dependence on both
the coupling constant and the cosmological constant, and compare with recent
scaling results in standard 2D quantum gravity.

\pagestyle{empty}

\end{abstract}
\end{titlepage}

\section{Introduction}

We have recently introduced in ref.[1] a new action for the gravitational
interaction, which we have called ``polytopic action", which  reduces to the
standard Einstein-Hilbert action at low energies and is bounded from below,
even though it is non local. Its basic variable is the two point geodesic
distance $ s(x,y)$, which is a well defined object in compact Riemannian
spaces as long as the two points are close enough, i.e. if the
geodesic segment connecting the two points does not cross the cut locus of any
of the points. But the main property of the polytopic action is that it is
bounded from below in a very natural way. This opens up the possibility that
the
quantum theory --as defined formally by the path integral over Riemannian
metrics-- is better behaved than the corresponding quantum version of the
Einstein-Hilbert action. This could shed light on some of the outstanding
problems in quantum gravity (cf, for example, ref.[2]). This is in
addition to the fact that our theory generalizes in a natural way to metric
spaces, thus providing a natural candidate for a fundamental theory of the sort
we sought for in ref.[3].

The purpose of the present work is to report on some preliminary results of a
systematic research in this direction. We have begun by exploring the simplest
setting, that is, the two dimensional situation in which only compact two
dimensional manifolds, that is Riemann surfaces, contribute to the partition
function.

But even in this case the problem is quite formidable, because our action is
not conformally invariant. We can topologically classify all compact two
dimensional Riemann surfaces $\cal M$ by a simple invariant, the Euler
characteristic $\chi (\cal M)$ which is given in terms of the genus $g$ (the
number of holes) by $\chi =2-2g$. The partition function will then be given by,
$$
Z(\kappa)= \sum_{g=0}^\infty \int {\cal D}g_{\alpha\beta}\exp
(-S_{poly}(\kappa)),
$$
where the integrals are taken at fixed genus, $S_{poly} (\kappa)$ is the
polytopic action and $\kappa$ is the coupling constant.

In the present work we
study the preceeding formula in the simplest cases of low genus, that is $g=0$
(the topology of the sphere) and $g=1$ (the topology of the torus). We can
parametrize each two dimensional metric using the conformal gauge (cf, for
example, ref.[4])

$$
g_{\alpha\beta}(x)= \exp(2\varphi) f_{\xi}\cdot\hat g_{\alpha\beta}(m),
$$
where $\exp(2\varphi)$ is the Weyl rescaling,
$f_\xi$ represents the action of the diffeomorphism group over the surface, and
$\hat g _{\alpha\beta}(x)$ is a fidutial metric, which can be taken in such a
way that its curvature is constant, $\hat R=+1$ for the sphere, $\hat R=0$ for
the torus. Note that in the higher genus case, which we are not going to
consider in the present work, it is $\hat R=-1$ and that in this case it also
depends on $2g-2$ complex moduli $m_a$.

This gauge fixing procedure implies the introduction of a
Faddeev-Popov ghost system $(b,c)$ of conformal weights $(2,-1)$. Furthermore
the polytopic action is not conformally invariant and this means that after
dividing by the volume of the two dimensional diffeomorphism group we still
have
to integrate over the conformal (Liouville) factor. Assuming the usual
arguments

by David, Distler and Kawai [5] (note, however, that these are not as
compelling
here as in the standard Einstein-Hilbert action, because the polytopic
Lagrangian is not Weyl invariant) the path integral that we
are dealing with is,

$$
Z=\int dm_a[{\cal D}\varphi][{\cal D}b][{\cal D}c]
\exp\left(-S_{poly}(e^{\gamma\varphi} \hat g(m_a)) -S_{ghost}(b,c)
-S_L(\varphi)\right)
\eqno(1.1)
$$
where  $S_{ghost}=(1/2\pi)\int (b\bar\partial c+\bar
b\partial \bar c)$ and the dressed Liouville action is given [5] by,

$$
S_L={1\over 8\pi}\int
d^2x\sqrt{\hat g}\left({\hat
g}^{\alpha\beta}\partial_\alpha\varphi\partial_\beta\varphi + {\mu\over
8\pi\gamma^2}e^{\gamma\varphi}+{Q\over 8\pi}\varphi \hat R\right), \eqno(1.2)
$$
where $\mu$ is the
cosmological constant, and $Q=2/\gamma +\gamma$, corresponding to a conformal
system with central charge $c=1+3Q^2$, where the numerical value of the
Liouville dressing, $\gamma$, is determined by the condition that
$\exp(\gamma\varphi)$ is a marginal perturbation , i.e. a $(1,1)$ conformal
tensor. The integral over the $2g-2$ moduli $dm_a$ is a finite dimensional
integral.

\section{The genus zero contribution}

Let us now compute the contribution to the integral in (1.1) coming from
surfaces with the topology of the sphere ($g=0$, $\chi=2$).

The polytopic action reads in the two dimensional case \cite{bb1},

$$
S_{poly}={1\over 2\pi}\int d\theta S_p[R(\theta)\cdot e]
\eqno(2.1)
$$
where $R(\theta)$ is a two dimensional rotation matrix,
$$
S_p[e] \equiv \int d^2x(e) S_p[e,x]
\eqno(2.2)
$$
and
$$S_p[e,x]\equiv {C\over \kappa^4}\Omega(P_1,P_2),
\eqno(2.3)
$$
where  $(e)= det(e_a^{\ \mu})$, $e_a^{\ \mu}$ is the zweibein, $C$ is a
dimensionless parameter, the gravitational coupling constant $\kappa^2\equiv
8\pi G$, the points $P_1$ and $P_2$ are associated to the point $x$ through the
exponential map corresponding to a given zweibein ${\vec e}_a=e_a^{\
\mu}\partial_\mu$ in $x$ , $P_a\equiv \exp(\kappa {\vec e}_a)$, and
$\Omega(P_a,P_b)$ is one half the square of the geodesic distance between the
points $P_a$ and $P_b$:

$$
\Omega(P_a,P_b)\equiv {1\over2}\left(\int_{P_a}^{P_b} ds\right)^2.
\eqno(2.4)
$$

In ref.[1] we have shown that the action (2.1) reduces in the weak
coupling regime, to the Einstein-Hilbert action,

$$
S_{poly}=\int d^2x(e)\left({6\over\kappa^2} - {1\over 2}R+\kappa^2 {19\over
36}(2R_{\mu\nu} R^{\mu\nu}+ 3 R_{\mu\nu\rho\sigma}R^{\mu\nu\rho\sigma}) +
O(\kappa^4)\right).
\eqno(2.5)
$$

It is, however, evident from the very definition (2.1), that the polytopic
action, being defined as a sum of distances, is positive definite; so that it
is
natural to think that the Euclidean path integral will be better defined than
in the standard case in which the Einstein-Hilbert action is unbounded from
below.

Although the polytopic action is not local and is difficult to compute even
classically (we need to control the geodesics in our manifold), we can easily
show that it is not conformally invariant, at least for constant rescalings. In
fact, when $g_{\alpha\beta}\rightarrow \exp(2\lambda)g_{\alpha\beta}$ this
means
that $e_a^{\ \mu}\rightarrow \exp(-\lambda)e_a^{\ \mu}$ and
since $\Omega(P_a,P_b)$ remains invariant, we have that
$$
S_{poly}\rightarrow e^{2\lambda}S_{poly}.\eqno(2.6)
$$

We can, of course, explore the consequences of the fact that the path integral
in (1.1) should depend only on the combination $g\equiv \exp(\gamma\varphi)\hat
g$, which means that it should have the ``fake" conformal invariance,
$$
\varphi\rightarrow\varphi-\sigma/\gamma,\ \ \ \ \hat g\rightarrow
e^\sigma\hat g.\eqno(2.7)
$$
This would imply in our case [5] that
$$
[{\cal D}_{e^{\sigma}{\hat g}}\varphi] [{\cal D}_{e^{\sigma}{\hat g}}b]
[{\cal D}_{e^{\sigma}{\hat g}}c]\exp\left(-S_{poly}(e^{\gamma\varphi}
e^\sigma\hat g) -S_{bc}(\varphi, e^\sigma\hat g)- S_L(\varphi, e^\sigma \hat
g)\right)=$$ $$
=[{\cal D}_{\hat g}\varphi] [{\cal D}_{\hat g}b]
 [{\cal D}_{\hat g}c]\exp\left(-S_{poly}(e^{\gamma\varphi}\hat g)-
S_{bc}(\varphi,\hat g) -S_L(\varphi,\hat g)\right), \eqno(2.8)
$$
that is, the conformal invariance of the gauge fixed system. But this conformal
invariance has no physical meaning specific for the polytopic action; it is
there for any action, be it Weyl invariant or not.

What is physically evident now is that we have to integrate over all
diffeomorphisms in the functional integral ${\cal D}\xi$ and not only over
those orthogonal to the conformal Killing vectors (CKV), ${\cal D}\xi^\bot$,
because our fundamental action is not Weyl invariant. This implies that when
the
gauge is fixed and the ghosts are introduced, the integral over diffeomorphisms
compensates all the Diff group without any remaining factor $vol(CKV)$.

As it is well known (cf, for example, ref.[4]) the group generated by the
conformal Killing vectors of the sphere, the M\"obius group, is isomorphic
to $SL(2,{\bf C})\sim O(3,1)$ and thus of infinite volume. This is the reason
why in a theory such as strings, both Weyl and Diff invariant, the zero, one
and two point functions in the sphere are zero. But, since this is not our
case,
let us proceed.

The ghost determinant in the sphere is a constant (cf ref.[6]), which we
include in the normalization of the measure. There are no moduli in the sphere,
so that no integrals over $dm_a$ are necessary, and our computation has
eventually reduced to a calculation of the single integral,
$$
Z_{g=0}=\int [{\cal D}\varphi]\exp\left(-S_{poly}(e^{\gamma\varphi}\hat g)-
S_L(\varphi)\right).\eqno(2.9)
$$

The first step is to compute the polytopic action for an arbitrary metric on
the sphere. the easiest thing is to start from the metric corresponding to a
round sphere of radius $R$ [7]:
$$
ds^2= R^2\left(\delta_{\mu\nu}+ {x_\mu x_\nu\over 1-r^2}\right)
dx^{\mu}dx^{\nu}
   ,
\eqno(2.10)
$$
where $\mu,\nu=1,2$, $r^2=x^2+y^2$ and each point on the sphere is given by the
cartesian coordinates $(x,y)$ of the equatorial plane of the sphere, thus it
corresponds to two points of the sphere except when $r^2=R^2$. An arbitrary
zweibein, $e_a^{\ \mu}$, at the point $(x,y)$ is given by,
$$
\vec e_1={\sqrt{1-r^2}\over Rr}\left((x\cos \theta+y\sin \theta)\vec
e_x +(y\cos\theta-x\sin\theta) \vec e_y \right)
$$
$$
\vec e_2={1\over Rr}\left((y\cos \theta-x\sin \theta)\vec e_x
-(x\cos\theta+y\sin\theta)\vec e_y \right),
 \eqno(2.11)
$$
where $\vec e_x=\partial/\partial x$ and $\vec e_y=\partial/\partial y$.

On the other hand, the general form of the geodesics for the metric (2.10) is
\cite{bb7},
$$
x=A\sin(\tau/ R) +B\cos(\tau/ R),\ \
y=C\sin(\tau/ R) +D\cos(\tau/ R), \eqno(2.12)
$$
where the parameter $\tau$ is affine, that is, normalized in such a way that
$g_{\mu\nu}(dx^\mu/ d\tau)(dx^\nu/ d\tau)=1$, when the constants
satisfy the condition
$$
A^2+B^2+C^2+D^2= 1-(AD-BC)^2. \eqno(2.13)
$$
Owing to rotational invariance, all points on the sphere are equivalent and we
can choose the zweibein around the north pole $(x=y=0)$, so that
$$
\vec e_1={1\over R}\left(\cos\theta\vec e_x+\sin\theta\vec e_y\right),\ \ \
\vec
e_2={1\over R}\left(-\sin\theta\vec e_x +\cos\theta\vec e_y\right). \eqno(2.14)
$$
The exponential mappings defining the points $P_1$ and $P_2$ give the
coordinates,
$$
P_1=(\cos\theta \sin R^{-1},\ \sin\theta\sin R^{-1}),\ \
P_2=(-\sin\theta\sin R^{-1},\ \cos\theta\sin R^{-1}). \eqno(2.15)
$$
The geodesic linking these two points is given by (2.12),
we can impose that $\tau=0$ at $P_1$, which fixes $B=\cos\theta\ \sin R^{-1}$,
$D=\sin\theta\ \sin R^{-1}$, and also that it should cross the point $P_2$ for
a
value of the affine parameter $s=\hat s$,
$$
A\sin(\hat s/R)+\cos\theta\sin R^{-1}\cos(\hat s/R)=-\sin\theta\sin R^{-1},
$$
$$
C\sin(\hat s/R)+\sin\theta\sin R^{-1}\cos(\hat s/R)=\cos\theta\sin R^{-1}.
$$
Here we still have to impose the condition (2.13) which now reads,
$$
A^2+C^2+\sin^2R^{-1}= 1+\sin^2R^{-1}(A\sin\theta-C\cos\theta)^2.
$$
It is now straightforward to compute $\hat s$ and we find that
$$
\cos(\hat s/R)= \cos^2R^{-1}. \eqno(2.16)
$$
Performing now the sum indicated in (2.1) and (2.2) one finds,
$$
S_{poly}(\kappa,R)={4\pi R^4\over \kappa^4}\left(\arccos
\left(\cos^2\left(\kappa/ R\right)\right)\right)^2. \eqno(2.17)
$$
Comparing with (2.9) we see that $R^2=\exp(\gamma\varphi)$. We are not able to
repeat the computation for arbitrary Liouville field $\varphi$, but we can
integrate over the constant zero mode, that is, over all round spheres. This is
obviously a very rough approximation (we are neglecting spheres with bumps and
warps), but we believe it still captures the essence of both the very small and
the very large distances regimes. We remark that in this approximation the
kinetic part of the Liouville action does not contribute, whereas the linear
curvature term $\hat R\varphi$ gives $(1/\gamma R^2)\ln R$ which, when
exponentiated conspires with the measure $d\varphi= (1/R)dR$ to give $R^a$
(with
$a$ a given power) whereas the cosmological constant term $\exp
(\gamma\varphi)$
just gives $\exp(\mu^2 R^2) $. We are then left with,
$$
Z_{g=0}(\kappa,\mu)\simeq\int_0^\infty dR\ R^a \exp\left(-S_{poly}(\kappa,
R)-\mu^2R^2 \right).\eqno(2.18) $$

In the Figure 1 we have represented the partition function for one illustrative
value of $a$ as a function of the cosmological constant, $\mu$,
and the gravitational coupling constant, $\kappa$. Besides the obvious general
trend of decreasing $\mu$ a curious phenomenon is a saturation of the integral
as a function of $\kappa$: it quickly reaches a strong coupling regime in which
the integral remains essentially constant.
 It is worth to compare with the partition function for the case of the
Einstein-Hilbert action, which in the same approximation reads,
$$
Z_{g=0,EH}(\kappa,\mu)\simeq\int_0^\infty dR\ R^a\exp\left(-{C\over2\kappa^2}
-\mu^2R^2\right),
$$
which unlike (2.18) has a trivial dependence on the gravitational coupling
constant. Thus even in this simplest case the quantum behavior of the polytopic
action differs considerably from that of the Einstein-Hilbert action.

\section{The genus one contribution}
In this section we consider the contribution to the integral (1.1) of the
$g=1$ topology. We shall, thus, evaluate the polytopic action on a two
dimensional torus which, as we have remarked in the introduction, is conformal
to the flat torus. As it is well known, a flat torus can be defined as follows.
Given a point $(a_1,a_2)$ in ${\bf R}^2$ in Cartesian coordinates $x$ and $y$,
w
   e
define the vector $\vec a=a_1\vec e_x + a_2\vec e_y$, where $\vec e_x$ and
$\vec
e_y$ are orthonormal vectors. Let us now define the vectors $\vec e_u=\tau_1
\vec e_x+\tau_2\vec e_y$ and $\vec e_v=\vec e_x$, and define the following
equivalence relation $\Gamma$ among the points of ${\bf R}^2$: two points
$(a_1,a_2)$ and $(b_1,b_2)$ are equivalent if the corresponding vectors $\vec
a$
and $\vec b$ satisfy that $\vec a-\vec b= m\vec e_u+ n\vec e_v$ for some
integers $m$ and $n$. The flat torus is defined as ${\bf R}^2/\Gamma$. The
region spanned by the vectors $\vec e_u$ and $\vec e_v$ with origin at the
coordinate origin is called the ``fundamental cell" (f.c.), and any point in
${\bf R}^2$ has a representative in the f.c.. To find the representative of a
point $(a_1,a_2)$ we first write the vector $\vec a$ in terms of the new
vectors
$\vec e_u$ and $\vec e_v$, i.e. $\vec a=(a_2/\tau_2)\vec e_u +
(a_1-a_2\tau_1/\tau_2)\vec e_v$; then the representative of that point in the
f.c. are the components of the vector $\vec a_{fc}$ defined as,
$$
\vec a_{fc}=\vec a-\left[{a_2\over \tau_2}\right]\vec e_u-
\left[a_1-{a_2\tau_1\over \tau_2}\right] \vec e_v,
\eqno(3.1)
$$
where in this section, $[s]$ means the greatest integer not larger than
 $s$.

The metric of the torus, being conformal to the flat torus, takes the form
$ds^2=\exp (\varphi(x,y))(dx^2+dy^2)$ restricted to the f.c. It can be written
in terms of the adapted coordinates $u$ and $v$, as

$$
ds^2=e^{\varphi(u,v)}(\tau^2 du^2 + 2\tau_1 dudv+ dv^2), \eqno(3.2)
$$
where $\tau^2=\tau_1^2+\tau_2^2$ and $0\leq u,v < 1$. Latter on we shall take
$\varphi(u,v)= constant$, since as for the case of the sphere we shall only
integrate over the constant zero modes.

As we recalled in the previous section to construct the polytopic action we
fix at an arbitrary point $P=(x,y)$ of the torus, i.e. the f.c., an orthonormal
zweibein $\vec e_1$ and $\vec e_2$, and define the points $P_1$ and $P_2$ in
the
f.c. by the exponential map of the vectors $\kappa \vec e_1$ and $\kappa \vec
e_2$ respectively. Once $P_1$ and $P_2$ have been determined we compute
$\Omega(P_1,P_2)$ as one half the square of the geodesic distance between $P_1$
and $P_2$. The polytopic action can then be constructed by first averaging
$\Omega(P_1,P_2)$ over all zweibein rotations and then by integrating over the
f.c., using (3.2).

The points $P_1$ and $P_2$ are easily found as the components of the vectors
$\vec p_{(1)}\equiv \vec p+(\kappa \vec e_1)_{fc}$ and
$\vec p_{(2)}\equiv \vec p+(\kappa \vec e_2)_{fc}$, where $\vec p=x\vec e_x+
y\vec e_y$. Now the distance between these two points in the f.c. is not
necessarily the length of the vector $\vec p_{(1)}-\vec p_{(2)}$ because
the minimum length might be obtained by a straight line joining the points
through the boundaries (which are identified) of the f.c. The true distance
may be computed as follows. Let us define a ``centered cell" (c.c.) in ${\bf
R}^2$ as the region obtained by the translation of the f.c. by the vector
$-(1/2)(\vec e_u+\vec e_v)$, i.e. the center of the c.c. is the origin of the
f.c. Let $\vec a= \kappa (\vec e_1-\vec e_2)$, then the true distance between
$P_1$ and $P_2$ is the length of the vector $\vec a_{cc}$ of the c.c.,
$$
\vec a_{cc}=\vec a- \left[{a_2\over\tau_2}+{1\over2}\right]\vec e_u-
\left[a_1-{a_2\tau_1\over\tau_2}+{1\over2}\right]\vec e_v. \eqno(3.3)
$$
This is easily seen as follows, let us make a translation of the two points so
that one of them goes to the centre of the c.c., if the other point gets
also translated into the c.c. it is clear that the distance between the points
is the length of the vector joining them. But if the second point is not in the
c.c. the distance may be obtained as the length of the vector joining the
center
with the representative of that second point into the c.c. In each case eq.
(3.3
   )
follows. Note that we assume here that the torus is homogeneous , i.e.
$\varphi$
is constant so that all points are isomorphic under translations and the
geodesics are straight lines.

Let us choose the
zweibein as,
$$\vec e_1=e^{-\varphi}(\cos\theta\vec e_x+\sin\theta\vec e_y),\ \
\vec e_2=e^{-\varphi}(-\sin\theta\vec e_x+\cos\theta\vec e_y), \eqno(3.4)
$$
then the vector $\vec a$, has components,
$$ a_1=-\sqrt2\kappa
e^{-\varphi}\sin(\theta+\pifour),\ \  a_2=\sqrt2\kappa
e^{-\varphi}\cos(\theta+\pifour).\eqno(3.5)
$$
 It is convenient, also, to
introduce an angle $\alpha$, by $$ \tau_1=-\tau\cos(\alpha-\pifour),\ \ \
\tau_2=\tau\sin(\alpha-\pifour), \eqno(3.6)
$$
then we have the expressions,
$
a_1\tau_2-a_2\tau_1=\sqrt2\kappa\tau e^{-\varphi}\cos(\alpha+\theta)$ and
$a_1\tau_1+a_2\tau_2=\sqrt2\kappa e^{-\varphi}\sin(\theta+\theta)
$, which turn out to be useful in what follows.

The square of the distance between $P_1$ and $P_2$ is, after using the above
expressions,
$$
2\Omega(P_1,P_2)= (\vec a_{cc})^2= k^2
+\tau^2\left[{k\over \tau_2}\cos (\theta+\pifour)+{1\over2}\right]^2
+\left[{k\tau\over \tau_2}\cos (\alpha+\theta)+{1\over2}\right]^2
$$
$$
-2k\tau\sin(\alpha+\theta)\left[{k\over\tau_2}\cos(\theta+\pifour)+{1\over2}
\right]
+2k\sin(\theta+\pifour)\left[{k\tau\over\tau_2}\cos(\alpha +\theta)+
{1\over2}\right]
$$
$$
+2\tau_1\left[{k\over \tau_2}\cos (\theta+\pifour)+{1\over2}\right]
\left[{k\tau\over \tau_2}\cos (\alpha+\theta)+{1\over2}\right], \eqno(3.7)
$$
where $k=\sqrt{2}\kappa\exp(-\varphi)$.

Next we compute the angular average of that function (see also (2.1)),
$$
\left<2\Omega\right>\equiv {1\over2\pi}\int_0^{2\pi}2\Omega(P_1,P_2).
\eqno(3.8)
$$
Since this involves the average of integer values of periodic functions, let us
consider the average $\left<\sin(\alpha+\theta)[a\cos(\theta+\chi)+
{1\over2}]\right>$ for some arbitrary $a$ and $\chi$, in some detail.
We first write,
$$
\left<\sin(\alpha+\theta)\left[a\cos(\theta+\chi)+\half\right]\right>=
\cos(\alpha-\chi)\left<\sin\theta\left[a\cos\theta+\half\right]\right>
$$
$$
+\sin(\alpha-\chi)\left<\cos\theta\left[a\cos\theta+\half\right]\right>,
$$
and note that the first term vanishes, i.e.
$$
\left<\sin\theta\left[a\cos\theta+\half\right]\right>=0.\eqno(3.9)
$$
This is easily proven by changing $\theta$ by $\pi-\theta$ (change of angular
origin) and by using that $[-s]=-1+[s]$ for an arbitrary real $s$. The
computation of the second term is more subtle. First we note that $
\left<\cos\theta[a\cos\theta+\half]\right>$ is invariant under the changes of
$\theta$ by $\pi-\theta$ and of $\theta$ by $2\pi-\theta$, so that we need to
consider only $0\leq\theta\leq\pi/2$, i.e.
$$
\left<\cos\theta\left[a\cos\theta+\half\right]\right>={2\over\pi}
\int_0^{\pi/2}d\theta \cos\theta \left[a\cos\theta+\half\right].\eqno(3.10)
$$
Now for a given $\theta$ let $n$ be the integer $n=[a\cos\theta+{1\over2}]$,
this means that $n\leq a\cos\theta+{1\over2}< n+1$. The contribution to the
average (3.10) of the angular interval from
$\theta_{n+1/2}\equiv\arccos({n+1/2\over a})$ to
$\theta_{n-1/2}\equiv\arccos({n-1/2\over a})$ is given by $
\int_{\theta_{n+1/2}}^{\theta_{n-1/2}} d\theta n\cos\theta=
-n\sqrt{1-\left({n-1/2\over a}\right)^2}+ n
\sqrt{1-\left({n+1/2\over a}\right)^2}$.
For $\theta=0$, $[a\cos\theta+\half]=[a+\half]\equiv N$, and for
$\theta=\pi/2$,
$[a\cos\theta+\half]=0$, therefore $0\leq n\leq N-1$. For $n=0$ there is no
contribution to the average (3.9) and for $N$ it is,
$
\int_0^{\theta_{N-1/2}}d\theta N\cos\theta= -N\sqrt{1-\left({N-1/2\over
a}\right)^2}.
$
Finally adding all contributions from $n=0$ to $N$ and combining the terms
appropriately we get
$$
\left<\cos\theta\left[a\cos\theta+\half\right]\right>={2\over\pi}\sum_{n=1}^N
\sqrt{1-\left({n-1/2\over a}\right)^2}. \eqno(3.11)
$$
Another term which can be computed by a similar method is,
$$
\left<\left[a\cos(\theta+\chi)+\half\right]^2\right>= 2\sum_{n=1}^N
(n-\half)\arccos({n-1/2\over a}). \eqno(3.12)
$$
Using (3.11) and (3.12) in (3.8) we get the final result,
$$
\left<2\Omega\right>=k^2
+2\tau_1\left<\left[{k\over\tau_2}\cos\theta+{1\over2}\
   right]
\left[{k\tau\over\tau_2}\cos(\alpha+ \theta -\pifour) +{1\over2}\right]\right>
$$
$$
-{4k\tau_2\over\pi}
\sum_{n=1}^{\left[k/\tau_2+\half\right]}
\sqrt{1-\left({n-1/2\over k/\tau_2}\right)^2}-
{4k\tau_2\over\pi\tau}\sum_{n=1}^{\left[k\tau/\tau_2+\half\right]}
\sqrt{1-\left({n-1/2\over k\tau/\tau_2}\right)^2}\eqno(3.13)
$$
$$
+2\tau^2\sum_{n=1}^{\left[k/\tau_2+\half\right]} (n-\half)
\arccos\left({n-1/2\over k/\tau_2}\right)
+2\sum_{n=1}^{\left[k\tau/\tau_2+\half\right]} (n-\half)
\arccos\left({n-1/2\over k\tau/\tau_2}\right).
 $$
We have not been able to find an analytical expression for
the second term of (3.13).

The polytopic action, as defined in ref. [1] is given by integrating
$\left<\Omega \right>$ over the surface of the torus, using (3.2):
$$
S_{poly}(\kappa,\varphi) =\int_0^1\int_0^1 dudv
e^{2\varphi}\tau_2\left<\Omega\right>= e^{2\varphi}\tau_2\left<\Omega\right>.
\eqno(3.14) $$

In the partition function (1.1) for the torus
there are no integrals over $dm_a$  (as in the case of the sphere). Thus the
partition function is reduced to the integral (2.9) for $g=1$ and the ghost
contribution. Again if we restrict ourselves to the constant zero mode for the
Liouville field ($\varphi$ is just a constant), such integral is reduced to,
$$
Z_{g=1}(\kappa,\mu)\simeq\int {d^2\tau\over\tau_2^{\ 2}}\int_{-\infty}^{\infty}
d\varphi{|\eta(\tau)|^4\over\tau_2} \exp\left(-S_{poly}(\kappa,\varphi) -\mu^2
e^\varphi\right), \eqno(3.15)
$$
where $\eta(\tau)$ is Dedekind's function.

In Figure 2 we have represented this integral for the case of the rectangular
torus, i.e. $\tau_1=0$ (in which case the polytopic action (3.14) simplifies
considerabily  because the ghost contribution $|\eta|^4/
\tau_2$ is just a constant absorbed in the overall normalization) in terms of
the couplig constant $\kappa$ and the cosmological constant $\mu$. The
vanishing of the integral as $\kappa$ and $\mu$ increase is noted here. It
is worth to compare this with the partition function for the  Einstein-Hilbert
action (which is zero for the flat torus)
$$
Z_{g=1,EH}(\mu) \simeq\int_{-\infty}^\infty
d\varphi\exp\left( -\mu^2 e^\varphi\right),
$$
which is independent of $\kappa$. We
see that for large $\varphi$, i.e. at low energies, the partition functions
agree, as one may expect, however, at high energies they are completely
different. Thus the quantum behaviour of the two theories is quite different
even in these low dimensional simple geometries: $g=0$ and $g=1$. Which theory
is better for quantization  cannot be answered at present; the polytopic action
is clearly more complicated but it has the advantage that it may be implemented
in a metric space (this, in fact, was the motivation for its introduction [1])
which is not necessarily a smooth manifold.

\section{Conclusions and perspectives}
We have studied in this paper some of the consequences of substituting the
Einstein-Hilbert action (which is a topological invariant in two dimensions)
by the much more complicated polytopic action. Although we have not been able
to
obtain exact results, due to the very complicated behavior of the polytopic
action under Weyl rescalings, we can easily get some physical intuition on the
critical exponents of the theory.

Let us consider, for example, the string susceptibility $\Gamma$, which is
defined from the fixed area partition function at genus zero
$$
Z(A)=\int [{\cal
D}\varphi] e^{-S}\delta\left( \int e^{\gamma\varphi} \sqrt{\hat g} d^2\xi
-A\right). \eqno(4.1)
$$
As is well known the DDK ([5]) scaling argument was based in the fact that
under a constant shift
$$
\varphi\rightarrow \varphi + \rho/\gamma, \eqno(4.2)
$$
($\rho$ is an arbitrary parameter) the standard Einstein-Hilbert plus Liouville
action scales as $$
S\rightarrow S -Q(1-g)\rho/\gamma, \eqno(4.3)
$$
and leads ($g=0$ for the sphere) to the simple expression
$$
Z(A)=K A^{a/\gamma-1},
$$
which is in agreement with the result of Knizhnik, Polyakov and
Za\-mo\-lod\-chi\-kov (KPZ) [8] ($K$ is an arbitrary parameter).

In our case, under the same transformation (4.2)
$$S\rightarrow S_{(ghost+Liouville)}- Q\rho/\gamma- e^{\rho}S_{poly}.
\eqno(4.5)
$$
There are two regions in which (4.5) reduces to the simple behavior of (4.3):
at the weak coupling regime ($\kappa=0$), in which $S_{poly}\rightarrow 0$, and
at the strong coupling regime in which, in the framework of our
approximations, $S_{poly}$ also tends to zero. In between, the string
susceptibility $\Gamma$ gets positive corrections; if, for example, $S_{poly}$
i
   s
constant, $$
\Gamma= \Gamma_{KPZ}+ S_{poly}. \eqno(4.6)
$$
More work is clearly needed before being able to asses the physical
significance of our results and, in particular, the detailed structure of the
small area behavior of the partition function. We hope to report on this in the
near future.

\section{Acknowledgements}
We are grateful to J.L.F. Barb\'on and M.A.R. Osorio for many useful
conversations on this topic. This work has been partially supported by CICYT
projects.

\newpage
\section{Figure captions}
\ \ \ \ \ Fig.1.--The integral (2.18) for genus $g=0$ is represented in terms
of
$\kappa$ and $\mu$ for $a=1$ (different values of $a$ lead to similar results).

Fig.2.--The integral (3.15) for genus $g=1$ is represented in terms of $\kappa$
and $\mu$, for $\tau_1=0$  ($\tau_2=1=\tau$). To regulate the ultraviolet
divergence (which is also present for the Einstein-Hilbert action) a short
distance cut off $\exp(\varphi)= 0.01$ has been used. The qualitative behavior
does not change with this parameter.

\end{document}